\documentclass[conference]{IEEEtran}
\usepackage{cite}
\usepackage[utf8]{inputenc}
\usepackage{booktabs}       % professional-quality tables
\usepackage{amsfonts}       % blackboard math symbols
\usepackage{amsmath}
\usepackage{color, soul}
\usepackage{graphicx}
\graphicspath{ {./figures/} }
\usepackage{subfig}
\usepackage[sort&compress,numbers]{natbib}
\usepackage{hyperref}

\title{Evaluation of the criticality of \\ \textit{in vitro} neuronal networks:
\\ \vspace{2mm} \LARGE Toward an assessment of computational capacity}

\author{\IEEEauthorblockN{Kristine Heiney\IEEEauthorrefmark{1}\IEEEauthorrefmark{2}\IEEEauthorrefmark{4},
Vibeke Devold Valderhaug\IEEEauthorrefmark{3},
Ioanna Sandvig\IEEEauthorrefmark{3}, \\
Axel Sandvig\IEEEauthorrefmark{3},
Gunnar Tufte\IEEEauthorrefmark{2},
Hugo Lewi Hammer\IEEEauthorrefmark{1}, and
Stefano Nichele\IEEEauthorrefmark{1}}
\IEEEauthorblockA{\IEEEauthorrefmark{1}\textit{Department of Computer Science,
Oslo Metropolitan University, Oslo, Norway}}
\IEEEauthorblockA{\IEEEauthorrefmark{2}\textit{Department of Computer Science,
Norwegian University of Science and Technology,
Trondheim, Norway}}
\IEEEauthorblockA{\IEEEauthorrefmark{3}\textit{Department of Neuromedicine and Movement Science,
Norwegian University of Science and Technology,
Trondheim, Norway}}
\IEEEauthorblockA{\IEEEauthorrefmark{4}Email: kristine.heiney@oslomet.no}}

\begin{document}

\maketitle

\begin{abstract}
Novel computing hardwares are necessary to keep up with today's increasing demand for data storage and processing power.
In this research project, we turn to the brain for inspiration to develop novel computing substrates that are self-learning, scalable, energy-efficient, and fault-tolerant.
The overarching aim of this work is to develop computational models that are able to reproduce target behaviors observed in \textit{in vitro} neuronal networks.
These models will be ultimately be used to aid in the realization of these behaviors in a more engineerable substrate: an array of nanomagnets.
The target behaviors will be identified by analyzing electrophysiological recordings of the neuronal networks.
Preliminary analysis has been performed to identify when a network is in a critical state based on the size distribution of network-wide avalanches of activity, and the results of this analysis are reported here.
This classification of critical versus non-critical networks is valuable in identifying networks that can be expected to perform well on computational tasks, as criticality is widely considered to be the state in which a system is best suited for computation.
This type of analysis is expected to enable the identification of networks that are well-suited for computation and the classification of networks as perturbed or healthy.
\end{abstract}

\section{Introduction} \label{intro}

Current computing technology is based on the von Neumann architecture, in which tasks are performed sequentially and control, processing, and memory are each allocated to structurally distinct components.
With this architecture, conventional computers struggle to cope with the rising demand for data processing and storage.
Furthermore, although recent advancements in machine learning technology have conferred great advantages to our data handling capabilities, processing continues to be performed on conventional hardware that has no inherent learning capabilities and thus requires huge amounts of training data, computational time, and computing power.

To continue to fulfill the rapidly growing computing demands of the modern day, it will be necessary to develop novel physical computing architectures that are self-learning, scalable, energy-efficient, and fault-tolerant.
The use of self-organizing substrates showing an inherent capacity for information transmission, storage, and modification \cite{langton1990edgeofchaos} would bring computation into the physical domain, enabling improved efficiency through the direct exploitation of material and physical processes for computation \cite{Stepney2018computationalmatter, jensen2018spinice}.
Some key properties of self-organizing systems that make them well-suited for computational tasks include their lack of centralized control and their adaptive response to changes in their environment \cite{heylighen1999}.
Such systems are composed of many autonomous units that interact with each other and the environment through a set of local rules to give rise to organized emergent behaviors at a macroscopic scale.
This type of spontaneous pattern formation is fairly common in nature, and there has been recent interest in determining how to develop interaction rules to generate various desired emergent behaviors \cite{doursat2013morphogenetic}, including those geared toward computation.
In addition, it has been demonstrated that self-organizing substrates can be used as computational reservoirs by training a readout layer to map the output of the physical system to a target problem \cite{schrauwen2007overview}.

The brain is an excellent example of a self-organizing system; it shows a remarkable capacity for computation with very little energy consumption and no centralized control, and scientists and engineers have long looked to the structure and behavior of the brain for inspiration.
Neurons grown \textit{in vitro} self-organize into networks that show complex patterns of spiking activity, which can be analyzed to gain insight into the network's capacity for information storage and transmission.
This behavior indicates that \textit{in vitro} neuronal networks may serve as a suitable computational reservoir \cite{aaser2017towards} and could also provide insights into the characteristics and dynamics desired for more engineerable substrates.

The aim of the present research project is to construct computational models that are able to reproduce desired behaviors observed in electrophysiological data recorded from engineered neuronal networks.
These models will provide insight into the behavior of the neurons and enable us to reproduce it in other substrates.
The computational capabilities of the models and different physical substrates developed from the models will be explored and their dynamics characterized.
This work is part of a project entitled Self-Organizing Computational substRATES (SOCRATES) \cite{socratesweb},
which aims to take inspiration from the behavior of \textit{in vitro} neuronal networks toward the development of novel self-organizing computing hardwares based in nanomagnetic substrates.

In addition to providing an avenue for the development of novel computational hardwares, the developed models are also expected to provide insight into the functionality of neuronal networks in healthy and perturbed conditions, where typical perturbations include chemical manipulation or electrical stimulation.
The dynamics of perturbed neuronal networks will also be modeled using the developed framework and their computational capabilities and dynamics characterized.
On the basis of this modeling, strategies of interfacing with perturbed networks to recover their dynamics will be explored.
The behavior of perturbed networks and their capacity for recovery will also provide insight into the robustness of the computational capabilities of engineered self-organizing substrates against analogous damage or perturbation.

The remainder of this paper is organized as follows.
Section \ref{results} presents an analysis method that will be used in this research project to assess the criticality of \textit{in vitro} neuronal networks toward identifying networks that may be considered well-suited for computation.
Results from the preliminary analysis of electrophysiological data recorded from a neuronal network cultured on a microelectrode array (MEA) are presented and discussed in this section.
%This analysis is based on the evaluation of the size distribution of neuronal avalanches and can be used to assess the criticality of an \textit{in vitro} neuronal network at the time of recording toward identifying networks that may be considered well-suited for computation.
A brief overview of the long-term plan for this research project is then given in Section \ref{plan}.
Section \ref{concl} concludes the paper.

\section{Neuronal avalanche analysis} \label{results}
\subsection{Background}
\begin{figure}
    \centering
    \subfloat[\label{MEA}]{%
        \includegraphics[width = 0.3\textwidth]{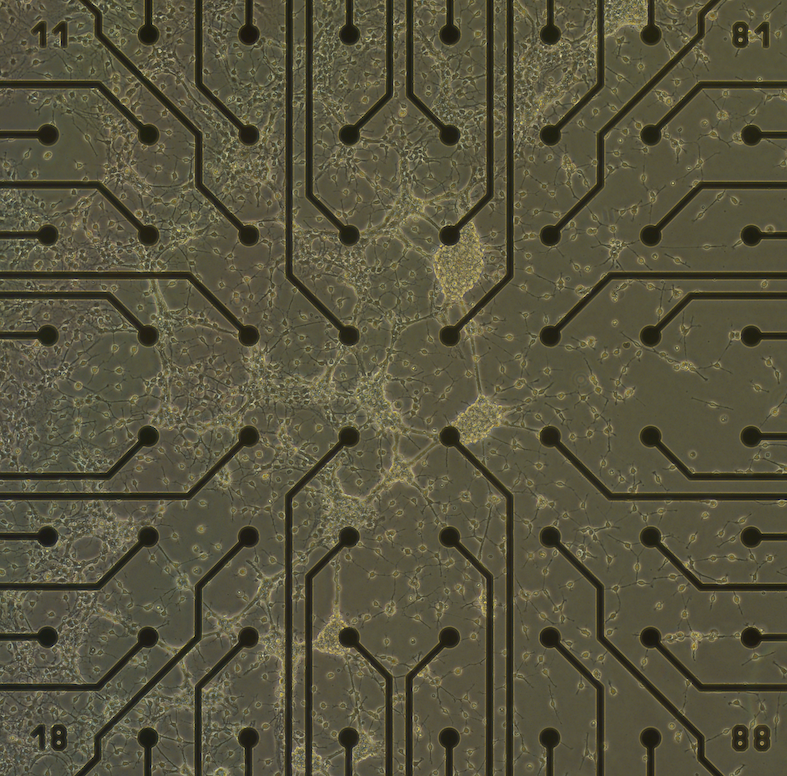}
    }\\ 
    \subfloat[\label{avalanche}]{%
        \includegraphics[width = 0.48\textwidth]{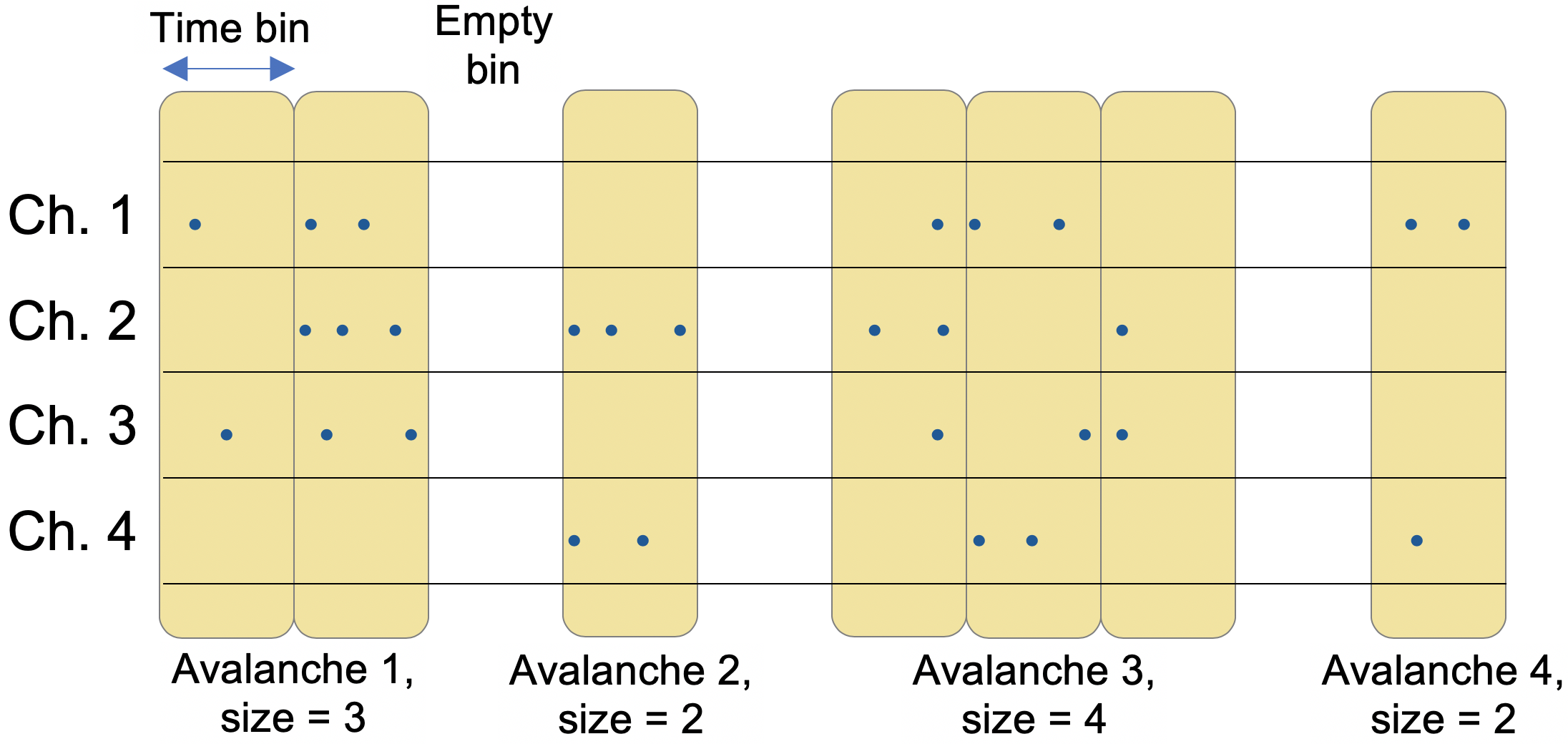}
    }
    \caption{(a) Microscope image of a neuronal network cultured on an MEA for illustrative purposes. Image taken by Ola Huse Ramstad. (b) Definition of a neuronal avalanche. Each dot represents a spike recorded by one of the electrodes (Ch. 1--4). A time bin is active when it contains at least one spike and empty when there are no spikes. An avalanche is defined as a sequence of consecutive active time bins preceded and followed by empty bins, and the size is the number of electrodes active during the avalanche.}
    \label{fig:avalancheMEA}
\end{figure}

An analysis method based on the size distribution of neuronal avalanches was applied to the analysis of an \textit{in vitro} neuronal network in this study (see Fig.~\ref{MEA} for an example of such an \textit{in vitro} network); this method is based on previous analysis performed on cortical networks \cite{pasquale2008, massobrio2015}.
The aim of this method is to determine whether a given neuronal network is in the critical state, which is presumed to be beneficial for the network in terms of its capacity to store information and perform computation.
A system in the critical state rests at the boundary between two qualitatively different types of behavior.
In the subcritical phase, a system shows highly ordered behavior characterized as static or oscillating between very few distinct states,
whereas in the supercritical or chaotic phase, the system shows highly unpredictable, essentially random behavior.
Near the transition point between these two regimes, the system is poised to effectively respond to a wide range of inputs as well as store and transmit information, making it ideal in terms of the capacity a system has for computation \cite{langton1990edgeofchaos}.

It has been proposed that the brain self-organizes into a critical state to optimize its computational properties; the foundations and theorized functional benefits of this behavior have been reviewed in recent articles \cite{shewplenz2013functional, hessegross2014}.
As first defined by Beggs and Plenz \cite{beggsplenz2003avalanches}, a neuronal avalanche is any number of consecutive time bins in which at least one spike is recorded, bounded before and after by time bins containing no activity, as shown in Fig.~\ref{avalanche}.
% It has been demonstrated both in cortical slice cultures \cite{beggsplenz2003avalanches} and dissociated cortical cultures \cite{pasquale2008, massobrio2015} that  [brief review of work that has been done so far on avalanches]
In their study, Beggs and Plenz \cite{beggsplenz2003avalanches} demonstrated that the size and duration of neuronal avalanches follows a power law, indicating that the propagation of activity in the cortex is in the critical state \cite{bak1987}.
It has been further demonstrated that criticality is established by a balance between excitation and inhibition and that cortical networks at criticality show greater dynamic range and information capacity and transmission than networks functioning outside of criticality \cite{shew2009, shew2011}.
Studies on the spontaneous activity of dissociated cortical networks have indicated that these networks tend to self-organize into the critical state over the course of their maturation, after first showing an early subcritical  followed by an intermediate supercritical phase \cite{tetzlaff2010developing, yada2017}.

As a long-term goal of the present research project, avalanche size distribution analysis will be applied to recordings obtained from different \textit{in vitro} neuronal networks to assess the criticality of the networks.
%The maturation of the networks will be observed and characterized for neurons cultured under normal conditions and those cultured with different perturbations toward characterizing the effect of the perturbations and classifying networks as healthy or perturbed.
Emerging network dynamics in healthy and perturbed conditions will be studied, characterized, and classified.
In the preliminary results presented here, the development of a single unperturbed neuronal network as it matures is reported.

To the authors' knowledge, this work represents the first time avalanche analysis has been applied to neurons derived from human induced pluripotent stem cells (iPSCs). Furthermore, this type of analysis has also not yet been applied to the characterization of the neuronal network dynamics in \textit{in vitro} disease models, which will be the focus of future work.

\subsection{Methods}
% Culture/recording methods kept short because this is a short paper.
The neuronal network assessed here was prepared as follows. Human iPSCs (ChiPSC18, Takara Bioscience) were reprogrammed using a protocol for midbrain dopaminergic neurons adapted from previous studies \cite{kirkeby2012culture, kirkeby2017culture, doi2014culture}.
Reprogramming was concluded on day 16, at which point the cells were left to mature.
The spontaneous electrophysiological activity of the network was recorded using a 60-electrode MEA together with the corresponding \textit{in vitro} recording system (MEA2100-System, Multi Channel Systems) and software (Multi Channel Experimenter, Multi Channel Systems).
Recordings of 6 min were taken starting after three weeks of maturation (day \textit{in vitro} (DIV) 21), starting from the date at which the reprogramming was concluded, and a total of 18 recordings taken over the period from DIV 21 to DIV 56 were analyzed.

Avalanches were detected according to the method described by Beggs and Plenz \cite{beggsplenz2003avalanches}.
Briefly, events were detected using thresholding on the data after applying a bandpass filter with a pass band of 300 Hz to 3 kHz \footnote{Code for spike detection is available at \url{https://github.com/SocratesNFR/MCSspikedetection}.}.
The spikes were then binned into time bins equal to the average inter-event interval (IEI), which is the time between events recorded across all electrodes, and avalanches were detected as any number of consecutive active time bins (bins containing at least one spike) bounded before and after by empty time bins.
The size of an avalanche is defined as the number of electrodes that were active during the avalanche.

A power law was then fitted to the avalanche size distribution data using a least-squares fitting followed by nonlinear regression with the result from the least-squares fitting for the initial parameter values.
This power law takes the form
\begin{equation}
    p(s) \propto s^{-\alpha},
\end{equation}
where $s$ is the avalanche size, $p(s)$ is the probability of an avalanche having size $s$, and $\alpha$ is the power of the fitted power law.
The fit was applied over the size range of $s = 2$ to $59$ electrodes, following previous works \cite{pasquale2008, massobrio2015}.
The goodness of fit was computed following Clauset et al.\ \cite{clauset2009powerlaw}.
Synthetic datasets were generated from the fitted distribution, and their Kolmogorov--Smirnov (KS) distances from the theoretical distribution were compared to the empirical KS distance.
The fitting was rejected if the fraction $p$ of synthetic KS distances that were lower than the empirical KS distance was greater than 0.1 ($p>0.1$) \footnote{Code for avalanche detection and goodness of fit evaluation is available at \url{https://github.com/SocratesNFR/avalanche}.}.

\subsection{Results and discussion}
The neuronal avalanche size distribution of a single \textit{in vitro} neuronal network was observed as the network matured.
In this preliminary work, no rigorous analysis was yet applied to classify network as super- or subcritical;
rather, only the goodness of fit of the size distribution to a power law was evaluated to assess whether the network was in a critical state during each analyzed recording.
Preliminary classification of non-critical cases was performed by visual inspection.

\begin{figure}
    \centering
    \subfloat[\label{pdfCritical}]{%
        \includegraphics[width = 0.45\textwidth]{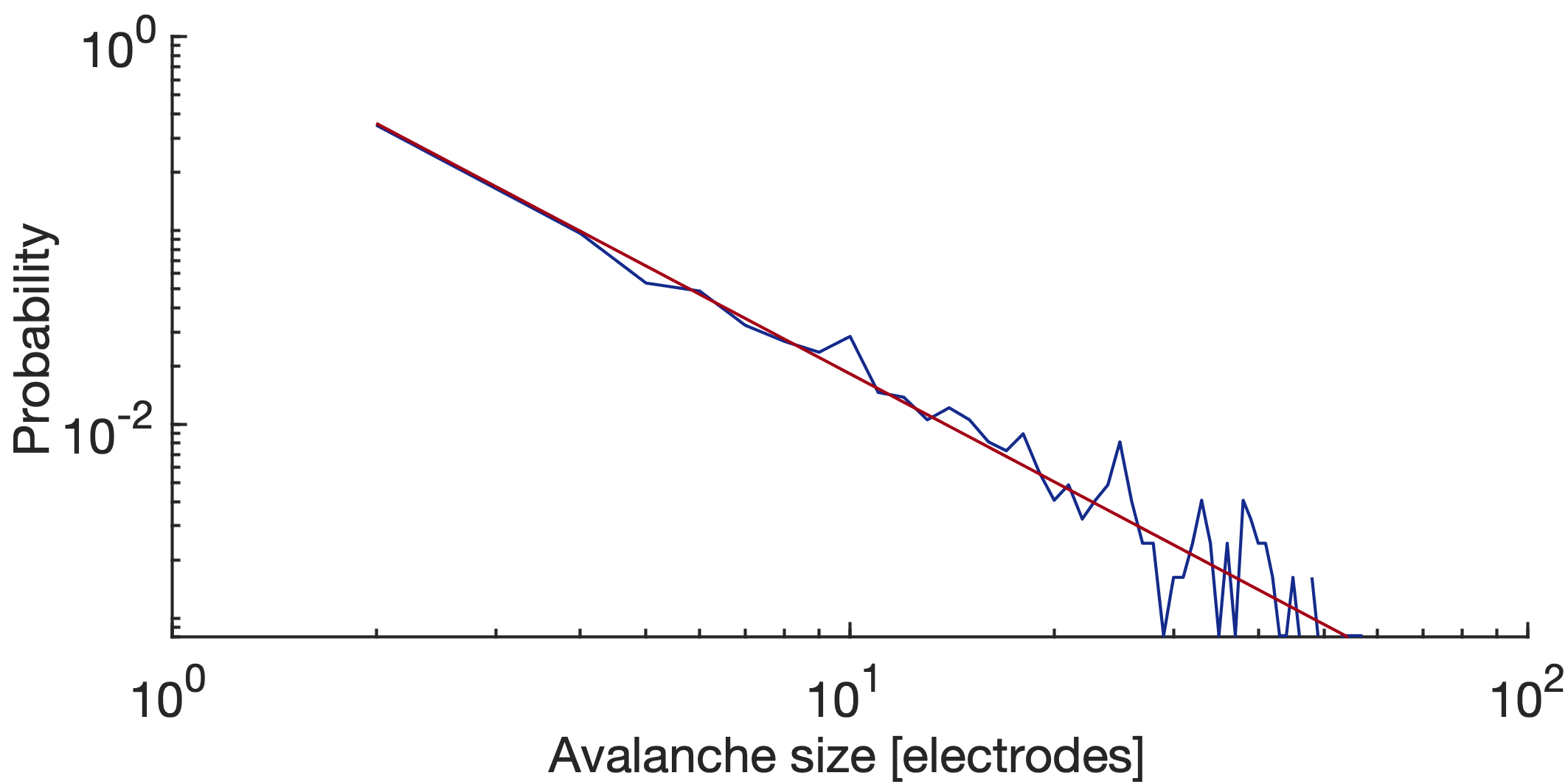}
    }\\
    \subfloat[\label{pdfSuper}]{%
        \includegraphics[width = 0.45\textwidth]{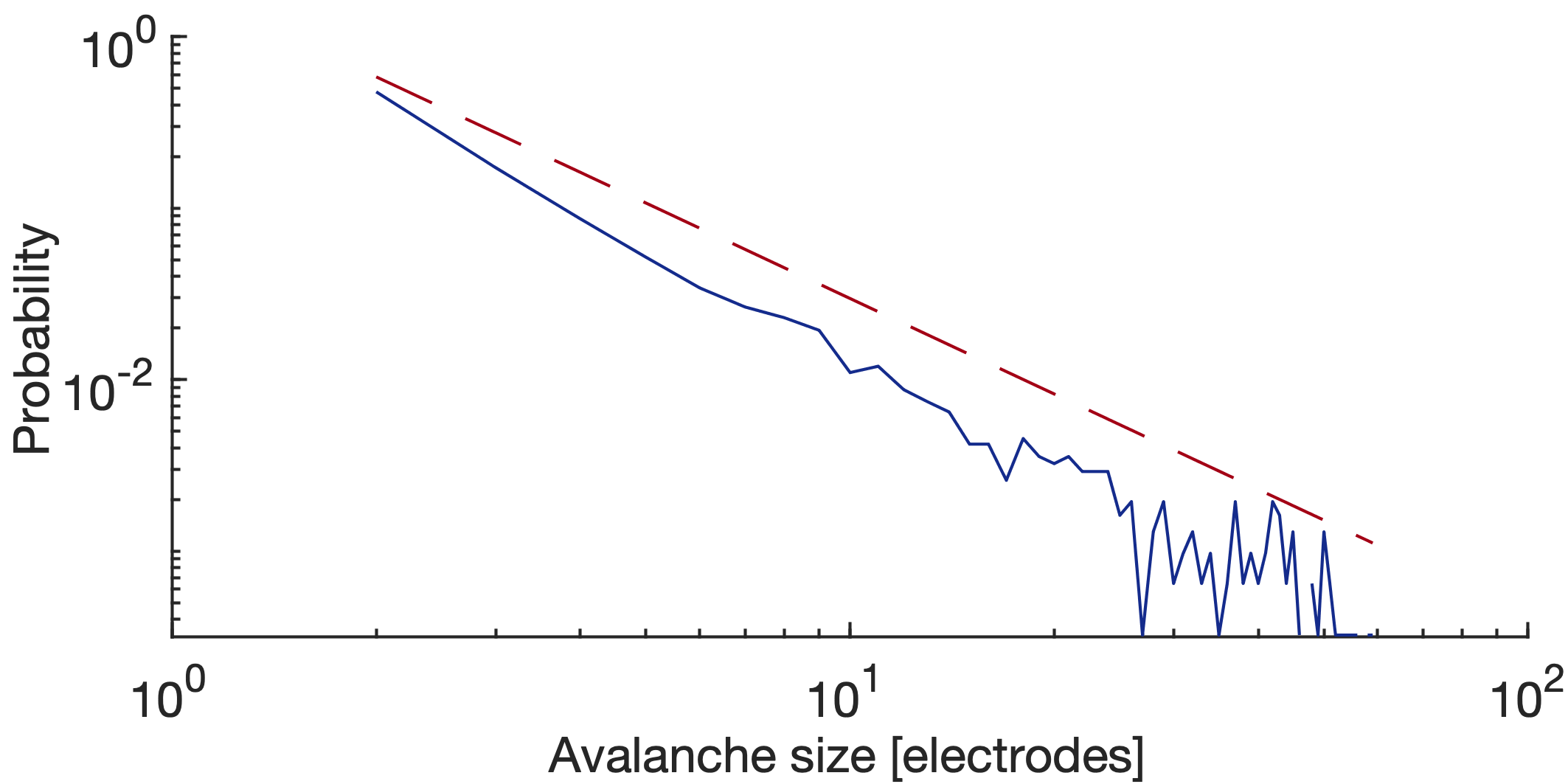}
    }\\
    \subfloat[\label{pdfSub}]{%
        \includegraphics[width = 0.45\textwidth]{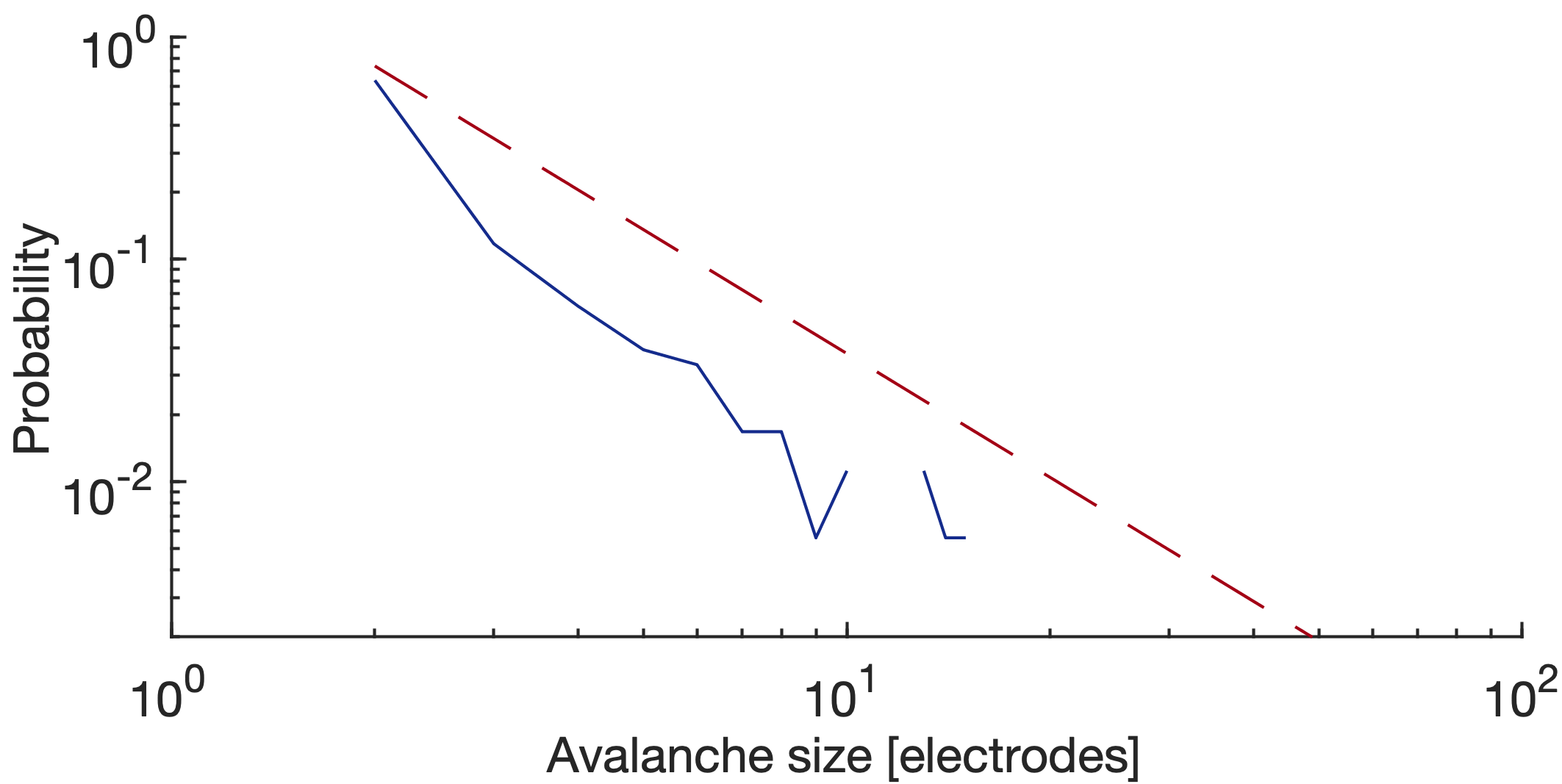}
    }
    \caption{Probability distribution functions for three representative cases. (a)~DIV 21: The fitting indicates the network is in a critical state, with $\alpha = 1.85$ ($p = 0.13$). The power-law fitting result is shown as a red line. (b)~DIV 51: The network appears to be in a supercritical state with a bimodal distribution. (c)~DIV 52: The network appears to be in a subcritical state with an exponential distribution. In (b) and (c), the dashed red lines correspond to $\alpha = 1.85$ for comparison with the distribution shown in (a).}
    \label{fig:pdfs}
\end{figure}

%Interestingly, the network did not show the same course of maturation as has been observed in previous studies on dissociated cortical networks (criticality reached after temporary periods of sub- and supercriticality).
The fitting results indicate that the network was already be in the critical state at DIV 21 (Fig.~\ref{pdfCritical}) and remained as such until DIV 42, with some brief deviations or periods of low activity.
Many of these deviations from criticality were during the early recording period, and the network was stably in the critical state between DIVs 36 and 42.
The mean power of the fitted power law distributions in the recordings where the network was in the critical state was $\alpha = 1.84 \pm 0.07$, which is higher than the value of $3/2$ reported by Beggs and Plenz \cite{beggsplenz2003avalanches}.
From DIV 44 until the final recording on DIV 56, the network was no longer in the critical state, and a preliminary visual assessment of the avalanche size distributions indicate that activity progressed to supercritical (Fig.~\ref{pdfSuper}) and finally subcritical (Fig.~\ref{pdfSub}) during the later recordings.
Supercritical behavior is characterized by a bimodal distribution, with an initially low slope in the log--log plot followed by a peak in the number of larger avalanches, whereas subcritical behavior is characterized by exponential decay, with few or no large avalanches; although no fittings were performed to rigorously assess whether the networks were in either of these states, the plots in Fig.~\ref{fig:pdfs} appear to be consistent with this type of behavior.

The observed network did not ultimately settle into a critical state in the considered timeframe, though it did pass through a period of relatively stable critical behavior.
The deviation from criticality is likely due to the different cell types that arise during the reprogramming of iPSCs, particularly as the proliferation of these cells causes the composition of the culture to change over time.
When differentiating iPSCs into a target cell type, it is impossible to avoid having other types of cells of the same lineage; for example, glial cells are inevitably present in iPSC-derived neuronal cultures.
This is different from networks that have been assessed in previous studies on self-organized criticality in neuronal networks, as these have focused solely on primary cortical networks, which can be prepared with greater homogeneity.
The heterogeneous and time-varying cellular composition likely produces changes to the signalling environment of the neurons, which may temporarily push the network away from criticality.
Additionally, the neurons assessed here were dopaminergic neurons, which are likely to show a different course of maturation in terms of criticality than cortical neurons.
It is possible that these types of networks show more complex oscillatory behaviors as they mature, or they may eventually settle into a critical state given enough time.
Further work is necessary to capture the expected time course of the development of the criticality of such networks.
%It is worth noting that this course of maturation is different from reported in previous papers on the development of cortical networks.
%However, these previous results also vary quite widely, with criticality being reached at around DIV 11 in one study \cite{yada2017} and DIV 58 in another \cite{tetzlaff2010developing};
%furthermore, as stated previously, the focus of this type of analysis has been on cortical networks, and this work represents the first time it has been applied to iPSC-derived dopaminergic neurons.

In future work, this analytical framework will be applied to recordings from additional unperturbed networks to gain a better understanding of the typical progression of networks as they mature \textit{in vitro}.
In cases where the data do not follow a power law distribution, further analysis will be performed to classify the state as sub- or supercritical;
these cases are known to show exponential and bimodal size distributions, respectively, as described previously.
This analysis will also be applied to networks that have been perturbed chemically or electrically to characterize how their behavior deviates from that of unperturbed networks and gain insights into how the perturbation may interrupt normal function.

\section{Plan for future research} \label{plan}
This work represents a first step in a larger research project, which will be described briefly here.
The plan for this research project is divided into four stages.
In the first stage, a data analysis framework will be developed, with the avalanche analysis method described here constituting a crucial part of this framework.
The framework involves methods of extracting meaningful features from electrophysiological data recorded from \textit{in vitro} neuronal networks.
Such features include conventional parameters considered in electrophysiological data analysis, such as the mean firing rate, as well as more complex measures, such as entropy and measures of connectivity.
The connectivity of the engineered networks will also be modeled using graph theory approaches.
The avalanche method presented here represents a useful tool for classifying networks as critical or non-critical.
Other methods of classification and clustering of networks will also be explored.

The second stage of the project involves the construction of computing models, such as cellular automata (CAs), random Boolean networks (RBNs), and recurrent neural networks (RNNs), that show behavior similar to that of the neuronal networks \cite{nichele2017deeplearningCA, nichele2017reservoirCA}.
The data analysis framework developed in the first phase will be used as a method of capturing the target behavior to be reproduced in the models,
and this framework will be continually refined as we improve our understanding of the important aspects of neuronal behavior that contribute to their computational capabilities.
These computing models are developed using evolutionary algorithms with appropriate fitness functions defined on the basis of the target behavior.
Important features of the models,
such as their input and output mappings and number of states,
will be explored,
and the dynamics of the models will be characterized.

The third phase involves the use of the developed models and the \textit{in vitro} neuronal networks as reservoirs to perform computational and classification tasks as a proof-of-concept using reservoir computing.
The models from the second stage will be refined based on their performance as computing reservoirs.

The final stage consists of the exploration of the application of the models developed in the second stage to the study of engineered neuronal networks under perturbed conditions mimicking pathologies related to the central nervous system (CNS).
Networks that have had their synaptic function perturbed will be modeled and analyzed using the developed methods to characterize how their behavior differs from that of unperturbed networks.
Methods of interfacing with the perturbed networks to restore their dynamics to the unperturbed state will then be explored.

\section{Conclusion} \label{concl}
The aim of this research project is to extract meaningful behaviors and features from electrophysiological data recorded from \textit{in vitro} neuronal networks and construct models that reproduce these behaviors toward the eventual realization of novel computing substrates based in nanomagnetic materials.
This paper reported the application of an avalanche size distribution analysis to electrophysiological data, representing a first step in the development of an analytical framework to extract target behaviors from such data.
The preliminary results reported here demonstrate emerging behavior that does not settle into criticality within the investigated time frame;
further work is needed to better characterize the time course of the development of criticality in the networks studied in this work and characterize how this affects the network's suitability for computation.

With this type of analysis, it can be determined if a network is in a critical state, which gives an indication of its suitability for use in computational tasks.
In addition to the computational applications of this analysis, it is also expected to be useful in distinguishing healthy and perturbed networks and to provide insight into how different diseases affect neuronal connectivity and communication, which will be the target of future work.

\section*{Acknowledgements}
This work was conducted as part of the SOCRATES project, which is partially funded by the Norwegian Research Council (NFR) through their IKTPLUSS research and innovation action on information and communication technologies under the project agreement 270961.

\bibliographystyle{ieeetran}
\bibliography{bib}

\end{document}